\documentclass[journal=jacsat,manuscript=article]{achemso}

\usepackage{chemformula} 
\usepackage[T1]{fontenc} 
\usepackage{amssymb}
\usepackage{xcolor}

\author{L. Ben Ltaief}
\affiliation{Department of Physics and Astronomy, Aarhus University, 8000 Aarhus C, Denmark}
\author{M. Shcherbinin}
\affiliation{Department of Physics and Astronomy, Aarhus University, 8000 Aarhus C, Denmark}
\author{S. Mandal}
\affiliation{Indian Institute of Science Education and Research, Pune 411008, India}
\author{S. R. Krishnan}
\affiliation{Department of Physics, Indian Institute of Technology, Madras, Chennai 600 036, India}
\author{A. C. LaForge}
\affiliation{Department of Physics, University of Connecticut, Storrs, Connecticut, 06269, USA}
\author{R. Richter}
\affiliation{Elettra-Sincrotrone Trieste, 34149 Basovizza, Trieste, Italy}
\author{S. Turchini}
\affiliation{Istituto Struttura della Materia-CNR (ISM-CNR), 00133 Roma, Italy}
\author{N. Zema}
\affiliation{Istituto Struttura della Materia-CNR (ISM-CNR), 00133 Roma, Italy}
\author{T. Pfeifer}
\affiliation{Max-Planck-Institut f{\"u}r Kernphysik, 69117 Heidelberg, Germany}
\author{E. Fasshauer}
\affiliation{Department of Physics and Astronomy, Aarhus University, 8000 Aarhus C, Denmark}
\author{N. Sisourat}
\affiliation{Sorbonne Universit\'e, CNRS, Laboratoire de Chimie Physique Mati\`ere et Rayonnement, UMR 7614, F-75005 Paris, France}
\author{M. Mudrich}
\email{mudrich@phys.au.dk}
\affiliation{Department of Physics and Astronomy, Aarhus University, 8000 Aarhus C, Denmark}

\title[Long-range charge exchange]{Charge-exchange dominates long-range interatomic Coulombic decay of excited metal-doped He nanodroplets}

\abbreviations{IR,NMR,UV}

\begin{document}
	

	\begin{abstract}
		Atoms and molecules attached to rare gas clusters are ionized by an interatomic autoionization process traditionally termed `Penning ionization' when the host cluster is resonantly excited. Here we analyze this process in the light of the interatomic Coulombic decay (ICD) mechanism, which usually contains a contribution from charge exchange at short interatomic distance, and one from virtual photon transfer at large interatomic distance. For helium (He) nanodroplets doped with alkali metal atoms (Li, Rb), we show that long-range and short-range contributions to the interatomic autoionization can be clearly distinguished by detecting electrons and ions in coincidence. Surprisingly, \textit{ab initio} calculations show that even for alkali metal atoms floating in dimples at large distance from the nanodroplet surface, autoionization is largely dominated by charge exchange ICD. Furthermore, the measured electron spectra manifest ultrafast internal relaxation of the droplet into mainly the 1s2s$^1$S state and partially into the metastable 1s2s$^3$S state. 
        \end{abstract}
      
Interatomic decay processes have recently been found to play a crucial role in the interaction of biological matter with energetic radiation. Both  free radicals and low-energy electrons produced by ICD processes can induce irreparable damage of the genome (double strand breaks in DNA) causing cancer or cell death~\cite{Sanche:2002,Trinter:2014,Gokhberg:2014}. Upon electronic excitation, weakly bound systems such as van der Waals or hydrogen bonded complexes and clusters can relax by interatomic autoionization if the excited state energy exceeds their adiabatic ionization energy. In the case of rare gas clusters doped with atomic or molecular impurities, this process has traditionally been termed Penning ionization~\cite{Kamke:1986,Froechtenicht:1996,Scheidemann:1997,Ren:2007,Wang:2008,Lan:2011,Buchta:2013}, in analogy to the collisional autoionization occurring in crossed atomic beams involving excited atoms, mostly rare gases prepared in metastable excited states~\cite{Siska:1993}.  This process is mainly driven by charge exchange between two interacting atoms or molecules which come so close to one another that their valence orbitals overlap. However, already in the early days of systematic Penning ionization studies, it was realized that the autoionization rate contains a second contribution describing energy transfer in the form of a virtual photon exchange~\cite{Katsuura:1965}. 

Since the seminal work by L. Cederbaum in 1997, such non-local autoionization processes involving two or more atomic or molecular centers have been formulated in the theoretical framework termed interatomic/intermolecular Coulombic decay (ICD)~\cite{Cederbaum:1997}. This approach mainly refers to the autoionization of weakly bound systems that are inner-shell excited by energetic photons or electrons, rather than to thermal collisions of metastable atoms. In such inner-shell excited complexes, the virtual photon transfer mechanism or direct ICD process mediated by energy transfer often dominates over the decay by charge exchange~\cite{Hergenhahn:2011,Jahnke:2015}. Direct ICD relies on the excited state being coupled to the ground state by an electric dipole-allowed transition and its decay rate scales as $\propto R^{-6}$ with the interparticle distance $R$, in contrast to the exponential scaling $\propto \exp\left( -R/a\right)$ of the charge exchange ICD term, where parameter $a$ depends on the spatial extension of the involved orbitals~\cite{Jahnke:2007,Jahnke:2015}. Therefore, the ICD by energy transfer occurs at rather large interatomic distances~\cite{SisouratNatPhys:2010}. However, charge exchange can significantly contribute to the ICD rate of excited dimers, especially at short interatomic distances where the valence electron orbitals overlap thereby enhancing the decay rate by orders of magnitude~\cite{Averbukh:2004,Kuleff:2010}. Thus, the autoionization of electronically excited clusters due to interatomic electronic couplings, traditionally named Penning ionization, can be regarded as one member of the family of ICD processes, and we argue that both terms are equally appropriate in the case presented here.

In experiments, the contributions to the non-local autoionization rate by charge and energy exchange are usually indistinguishable, since the final products (electrons and ions) are the same. However, T. Jahnke and co-workers have recently shown experimentally that the two contributions can actually be disentangled for the cases of Ne$_2$~\cite{Jahnke:2007} and HeNe~\cite{Sann:2017} dimers by inferring the interatomic distance where ICD takes place. In the current work, we present a method to discern the long and short-range contributions to ICD for a very different system -- helium nanodroplets doped with alkali metal atoms. This method relies on the property of clusters and nanodroplets to solvate ions, provided they are formed with low kinetic energy to prevent their detaching from the cluster~\cite{Vangerow:2015,Vangerow:2016}. We find that ICD occurring at large dopant-helium distances ($R\gtrsim 5$~\AA) strongly dominates over ICD at short range ($R\lesssim 5$~\AA), where the atomic energy levels are measurably shifted. Nevertheless, \textit{ab initio} calculations show that ICD proceeds nearly exclusively by charge exchange for all experimentally relevant distances. Furthermore, detailed insights into the internal relaxation of the excited He nanodroplets prior to ICD are gained. 

\begin{figure}
	\center
	\includegraphics[width=1.0\columnwidth]{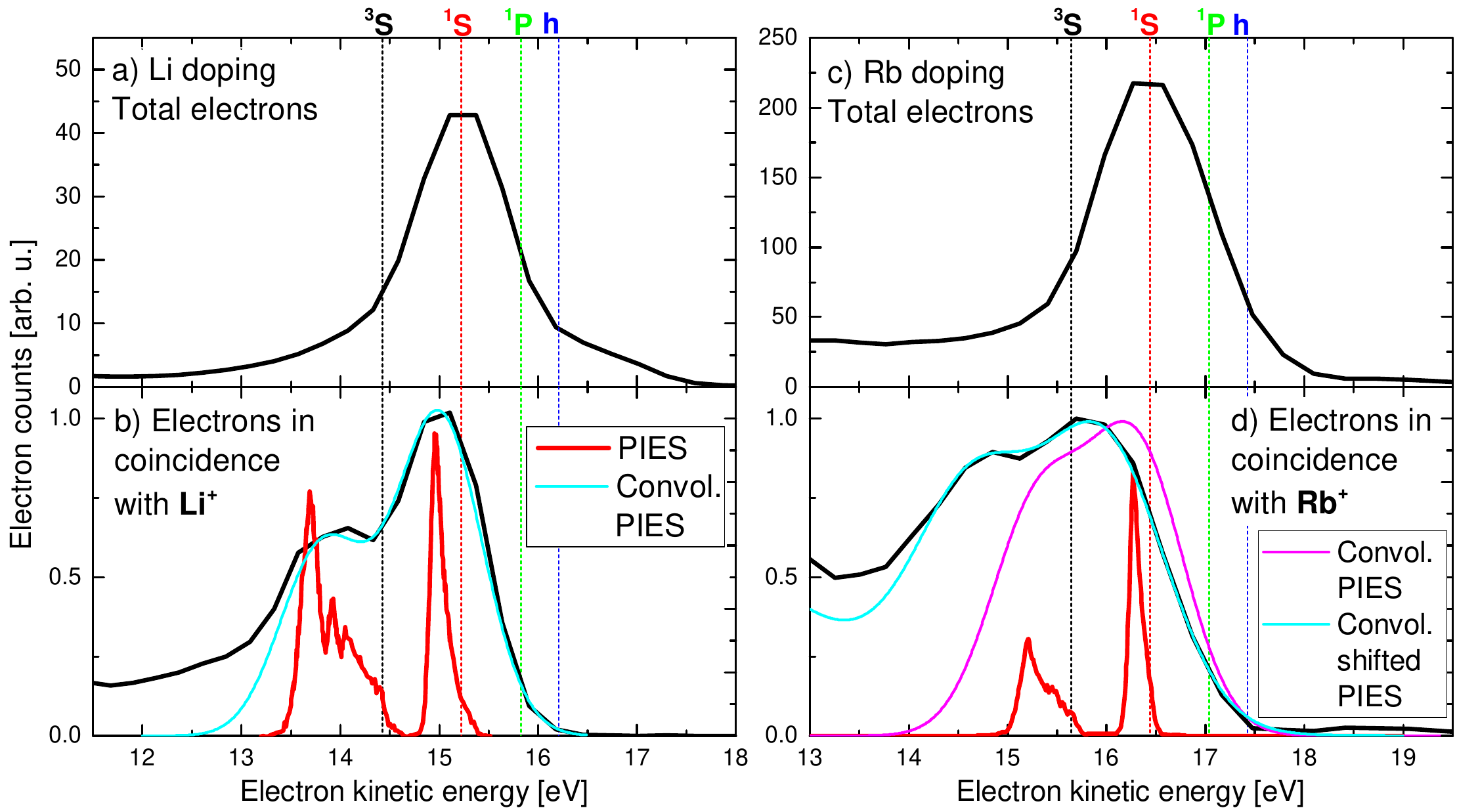}\caption{\label{fig1} Electron spectra recorded for He nanodroplets doped with Li atoms [a) and b)] and with Rb atoms [c) and d)] at a photon energy $h\nu=21.6$~eV. a) and c): Spectra of all electrons emitted from the doped He nanodroplet. b) and d): Spectra recorded in coincidence with Li$^+$ and Rb$^+$ dopant ions. The spectra in a) and b) as well as those in c) and d) are normalized to the same scaling factors to preserve their relative intensities, respectively. The red lines in b) and d) show Penning ionization electron spectra (PIES) measured in crossed atomic beams~\cite{Ruf:1987}. The light blue lines show these PIES for adjusted relative contributions of $^1$S and $^3$S components convoluted with the instrument function of the spectrometer (and slightly shifted in the case of Rb). The dashed vertical lines indicate expected electron energies based on the atomic term values and ionization potentials.}
\end{figure}

To perform these experiments, a beam of He nanodroplets containing on average $\bar{N}_\mathrm{He} =2.3\times 10^4$ He atoms per droplet is doped with alkali metal atoms M (Li, Rb) and irradiated by extreme-ultraviolet (EUV) light generated by a synchrotron light source. The EUV light is tuned to the strongest absorption band of He nanodroplets correlating to the 1s2p$^1$P state of He at a photon energy $h\nu = 21.6$~eV and to higher-lying bands~\cite{Joppien:1993}. Subsequently, the doped droplets autoionize in the reaction $(\mathrm{He})^*_N + \mathrm{M} \rightarrow \mathrm{He}_N + \mathrm{M}^+ + e^-$. The electrons $e^-$ and ions M$^+$ are detected in coincidence using a velocity-map imaging (VMI) spectrometer to measure the electron kinetic energy $K_e$~\cite{Eppink:1997,Buchta:2013}.  

Fig. 1 shows the electron spectra measured for He nanodroplets doped with Li atoms [panels a) and b)] and with Rb atoms [panels c) and d)] at $h\nu = 21.6$~eV. These spectra recorded for different average droplet sizes ($\bar{N}=2500$ and $2\times 10^6$) are essentially identical to those in Fig. 1. The top panels, a) and c), show the spectra recorded when measuring all electrons emitted from the droplets, while panels b) and d) show the same measurements taken in coincidence with Li$^+$ and Rb$^+$, respectively. The vertical dashed lines illustrate the electron energies $K_e=E_\mathrm{He}-E_i$ evaluated for the level energies $E_\mathrm{He}$ of He atoms in the excited states 1s2s$^3$S, 1s2s$^1$S, and 1s2p$^1$P assuming relaxation of the droplet excited states to atomic levels. The ionization energies of Li and Rb dopants are taken as the atomic values $E_i=5.39$~eV and $4.18$~eV, respectively. The vertical line marked by $h\nu$ depicts the electron energy $K_e=h\nu - E_i$ one would measure if ICD occurred directly from the resonantly excited state of the He droplet. 
However, we find that the excited He droplets initially relax into lower-lying levels near those of free He atoms prior to the ICD~\cite{Wang:2008,Buchta:2013}. The relaxation dynamics was recently directly mapped using EUV-UV pump-probe and a relaxation time of about 1~ps was found~\cite{MudrichRelaxation}. Electrons emitted by direct photoionization of the alkali atoms in the residual gas or on He nanodroplets, which would add to the signal at $K_e=h\nu - E_i$, do not contribute due to the low absorption cross section of the atoms.

One main finding here is that ICD predominantly occurs out of the 1s2s$^1$S state as seen from the sharp peaks in the total electrons spectra shown in the top panels a) and c) of Fig.~\ref{fig1}. The finite widths of the peaks are due to the limited resolution of our VMI spectrometer. This finding is in agreement with the time-resolved relaxation measurements which showed that most of the He droplets initially excited into the 1s2p$^1$P band end up as nearly unperturbed He atoms in the 1s2s$^1$S state residing in void bubbles inside the droplets or as free 1s2s$^1$S atoms ejected out of the droplets~\cite{MudrichRelaxation}. The fact that the peaks are nearly unshifted with respect to the atomic 1s2s$^1$S level indicates that the dopant-droplet autoionization occurs at large interatomic distance $R\gtrsim 5$~\AA~where the energy levels are only weakly perturbed. In contrast, in those events that generate both an electron and a free dopant ion [bottom panels b) and d)], the electrons are emitted at slightly lower kinetic energies. Note that we measure 30 and 200 times more electrons in total as compared to the number of electron-ion coincidence events involving Li$^+$ and Rb$^+$ ions, respectively. Assuming a detection efficiency for ions of 30\,\%, this results in ratios of total emitted electrons to coincident electrons of 9 and 60 for Li$^+$ and Rb$^+$ ions, respectively. This ratio is larger for Rb than for Li because the heavy Rb ion is less likely to detach from the droplet surface compared to the light Li ion. This was seen in similar experiments for alkali atoms and dimers attached to He nanodroplets, where the proportion of dopant ions ejected by ICD was lower for the heavier species when comparing to the ion signal due to double ICD of the dimers~\cite{LaForge:2019}. Detecting those ions that are fully solvated inside the He nanodroplets would require the ion spectrometer to be placed in line with the He droplet beam and the detector to be sensitive for large ion masses, as \textit{e.\,g.} in Ref.~\cite{Vangerow:2015}.

The double-peak structure present in all coincidence electron spectra [Figs.~\ref{fig1} b) and d)] resembles the Penning ionization electron spectrum (PIES) measured by M. W. Ruf and co-workers using crossed atomic beams, shown as red lines~\cite{Ruf:1987}. This PIES is characteristic for traditional Penning ionization occurring in collisions of He atoms prepared in both 1s2s$^1$S and $^3$S metastable states with groundstate atoms of a different species at thermal beam velocities. In such collisions, the interaction time of the colliding atoms is extremely short ($\ll 1$~ps) such that the autoionization is entirely dominated by the charge exchange mechanism which is active at short interatomic distance~\cite{Jahnke:2015}.

The light blue lines show the results of convolving the PIES with the instrument function of the VMI spectrometer. In the Rb case, the 1s2s$^3$S and 1s2s$^1$S peaks are shifted by $-0.65$ and $-0.33$~eV, respectively, in order to match the experimental and modeled spectra.
A Gaussian function centered around $9$~eV is added to account for an additional signal component due to double ICD~\cite{LaForge:2019}.
The electron energy shifts are likely due to elastic scattering of the emitted electron with the surrounding He atoms. We mention that the Penning electron spectra for molecules embedded in the droplet interior are even more strongly shifted and broadened~\cite{Shcherbinin:2018}. In addition, the amplitudes of the $^3$S component was scaled by factors 0.38 (Li) and 0.24 (Rb) relative to the ones of the $^1$S component to fit the experimental spectra.

\begin{figure}[htb]
	\center
	\includegraphics[width=0.6\columnwidth]{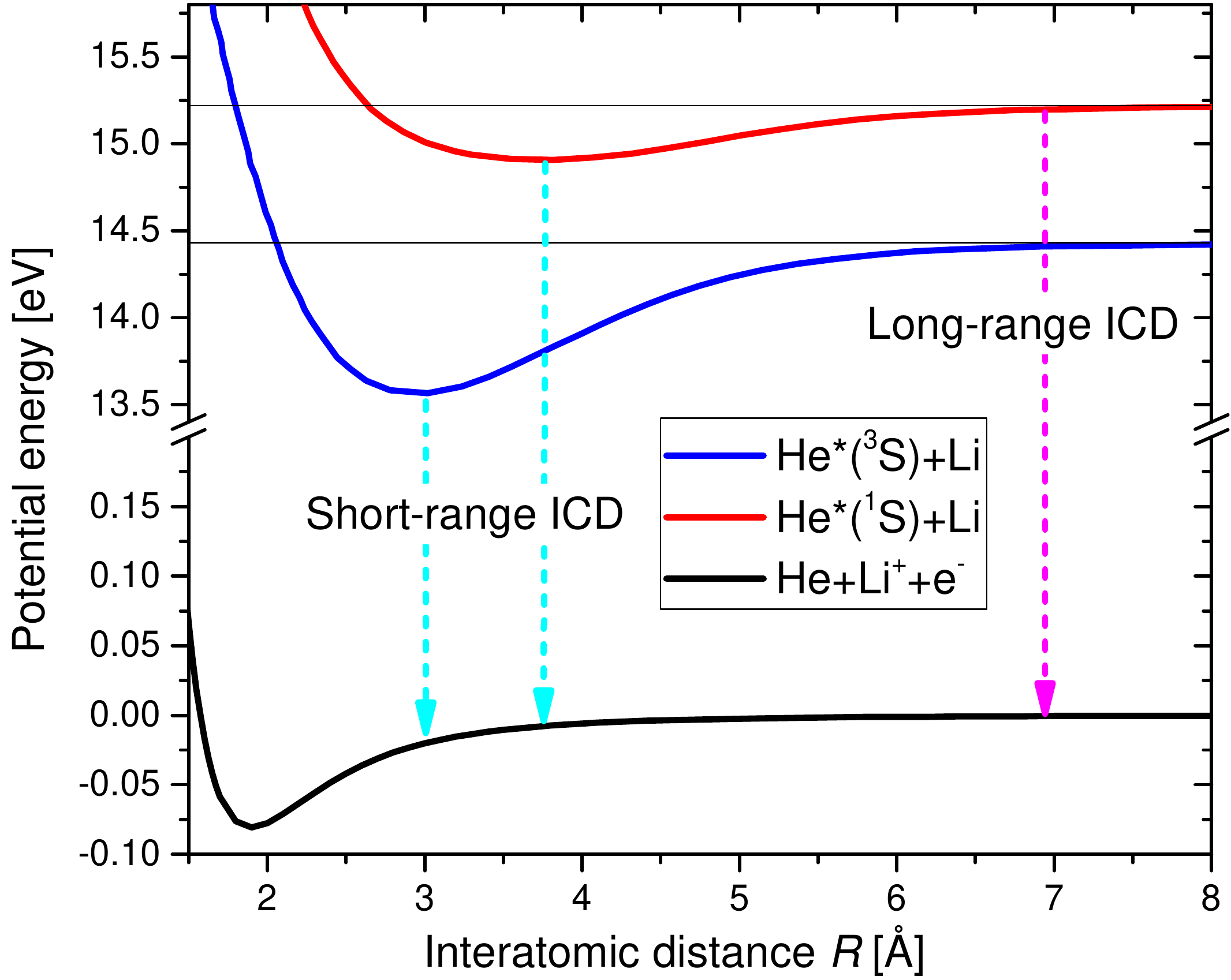}\caption{\label{fig2} Potential energy curves involved in the ICD process of Li and excited He$^\ast$, taken from~\cite{Movre:2000,Soldan:2001}. ICD taking place at large interatomic distance $R\gtrsim 5$~\AA~is represented by the pink vertical arrow. ICD occurring after the contraction of the He$^\ast$Li dimer to short distance $R=$2.5-4~\AA~is illustrated by the blue vertical arrows.}
\end{figure}	
Both the sharp peak in Fig.~\ref{fig1} a) and c) and the double-peak structures in Fig.~\ref{fig1} b) and d) can be understood when considering the potential energy curves of the excited and ionized HeLi dimer, shown in Fig.~\ref{fig2}. When the dopant atom is ionized at large interatomic separation $R$, the electron is emitted with an energy given by the atomic 1s2s$^1$S level. While for the free He atom the transition from this state to the 1s$^2\,^1$S ground state is dipole forbidden, this transition becomes partly allowed for the He$^\ast$Li dimer due to the breaking of the atomic symmetry (in the molecular symmetry the $^1\Sigma \rightarrow ^1\Sigma$-transition is allowed). 

When the He and Li atoms move towards each other along the attractive potential energy curves prior to decaying, the potential energy difference taken up by the electron as kinetic energy is reduced. At short distance $R\lesssim 5$~\AA,~the non-local autoionization process is dominated by charge exchange ICD. Therefore, it can occur both for the partly allowed 1s2s$^1$S state and for the 1s2s$^3$S state which remains metastable with respect to optical de-excitation due to the spin selection rule. Surprisingly, the double peak structure of the coincidence electron spectra (Fig.~\ref{fig1}) shows that a considerable fraction of excited He droplets relax into the 1s2s$^3$S state which requires a change of the electron spin multiplicity. Previously, fluorescence measurements had indicated that triplet states would only be populated by electron-ion recombination following He droplet autoionization occurring at $h\nu>23$~eV~\cite{Haeften:1997}. Thus, we find that droplet relaxation pertains not only to the atomic motion (bubble formation) and to the orbital angular momentum state of the excited-state electron (interband relaxation~\cite{Ziemkiewicz:2015}), but even to the spin degree of freedom. 

From the large ratio of unshifted electrons created by ICD at long range to those shifted to lower energies, we conclude that in most cases, ICD proceeds before the atoms have notably moved towards each other. But why is the coincidence detection of electrons and dopant ions selective to the short-range process, whereas the majority of electrons are created at long range? This is related to the peculiar property of He nanodroplets that excited electrons are repelled from their local He environment thereby leading to the formation of a void bubble, whereas ions tend to be attracted towards the He and form strongly bound snowball complexes with a dense shell of He atoms around the ion~\cite{Mueller:2009,Galli:2001}. As a result, an alkali atom ionized at the droplet surface is likely to sink into the droplet and thus eludes its detection~\cite{Vangerow:2015,Vangerow:2016}. Counter-intuitively, this happens when the atom is ionized at large distance (pink dashed arrow in Fig.~\ref{fig2}), because in that case the ion is created at the droplet surface nearly at rest. In contrast, when ionization occurs near the minimum of the potential energy well (blue dashed arrows in Fig.~\ref{fig2}) where the dopant atom has acquired a maximum of kinetic energy, the ion bounces back off the He and escapes from the droplet as the atom's kinetic energy is conserved in the ICD. This implies that the detected ratio of $^3$S versus $^1$S atoms is probably larger than the actual ratio of populations in these states. Since the He$^*$Li potential energy curve has a deeper well in the $^3$S state than in the $^1$S state, the kinetic energy and therefore the escape probability of the ions is enhanced for the $^3$S state. This is likely the reason why the $^3$S contribution is seen in the coincidence spectra but not in the total electron spectra. This picture is supported by the dopant ion kinetic energy distributions measured in ion VMI mode, see the Supplemental Material.

\begin{figure}
	\center
	\includegraphics[width=0.6\columnwidth]{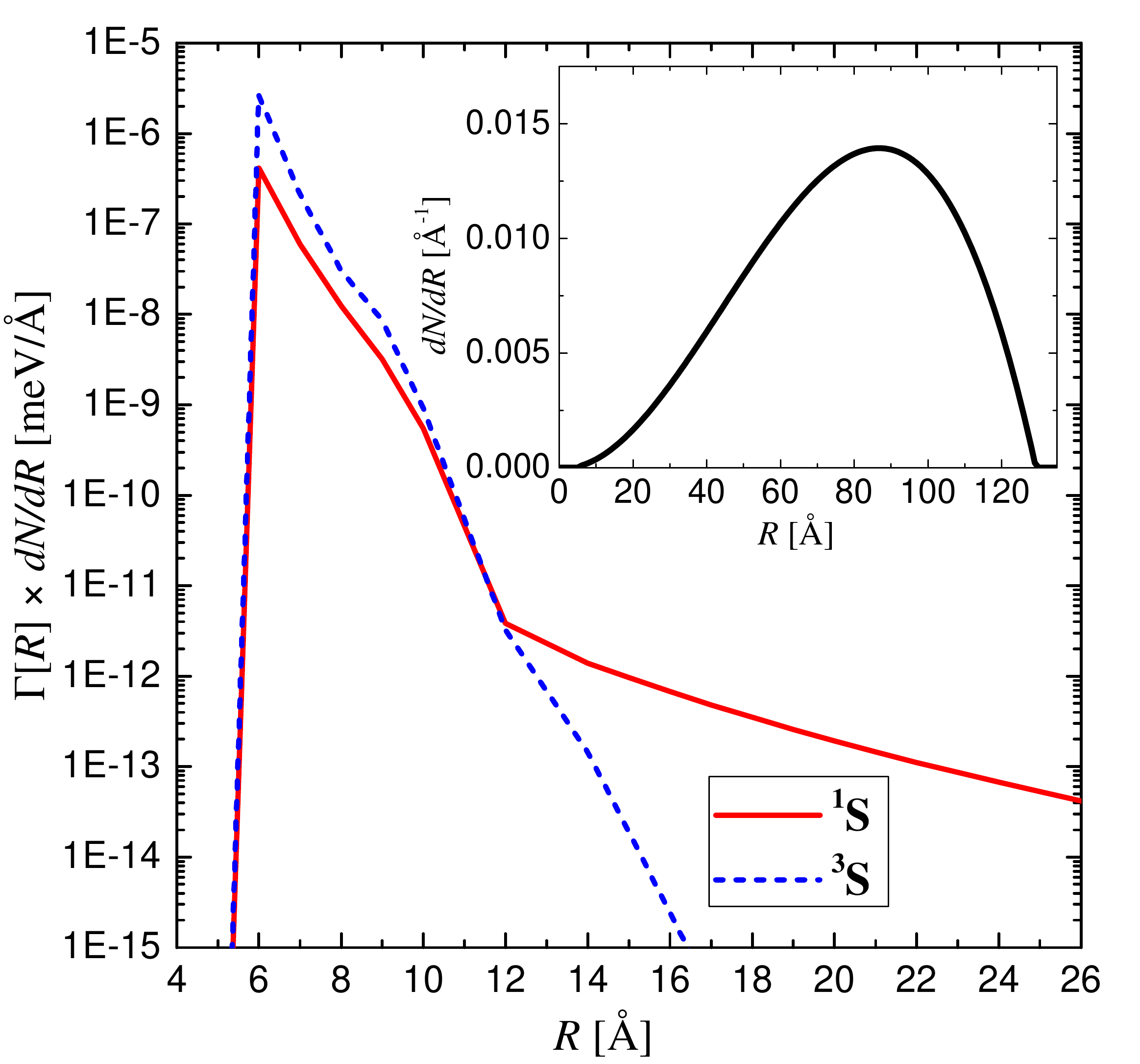}\caption{\label{fig5} Calculated average ICD rates for the He$^*$+Li reaction involving the two lowest excited states of He$^*$, 1s2s$^{1,3}\Sigma$, weighted by the number of He atoms present at distance $R$ from the Li atom, $N(R)$, in a He droplet of radius 63~\AA, see inset.}
\end{figure}
The clear distinction between ICD occurring at long and short ranges from the photoelectron spectra might lead one to conclude that in this way, the contributions from virtual photon and charge exchange ICD can be separated experimentally, as it was demonstrated for the neon dimer~\cite{Jahnke:2007}. To assess this conjecture, we have carried out \textit{ab initio} calculations of the ICD rates $\Gamma$ for the 1s2s$^1$S and 1s2s$^3$S states using the Fano-CI-Stieltjes method~\cite{Miteva:2017}. Computational details will be given in a future publication. The results show that this is not the case for the He$^*$Li dimer. In this system, charge exchange ICD dominates over direct ICD for interatomic distances ranging at least up to 12~\AA. In this range, $\Gamma$ follows the expected exponential scaling with $R$. Only for $R>12$~\AA~can $\Gamma$ be extrapolated by the power law $\propto R^{-8}$ for the $^1$S state, as direct ICD by virtual photon transfer is possible via electric quadrupole transition~\cite{Averbukh:2004}. 

Fig.~\ref{fig5} shows the expected ICD rates $\Gamma_D^{^1\mathrm{S},\,\,^3\mathrm{S}}=\Gamma^{^1\mathrm{S},\,\,^3\mathrm{S}}\times dN/dR$ weighted by the number of He atoms present in a He droplet at distance $R$ from the Li atom, $dN(R)/dR$, see the inset. For a spherical shape of the He nanodroplet, $dN(R)/dR$ is given by
\[
\frac{dN(R)}{dR}=\frac{2\pi R}{V_D} \left(\frac{R_D^2-R^2}{2R_\mathrm{Li}}+R-\frac{R_\mathrm{Li}}{2}\right)
\]
such that $\int dN(R)/dR \,dR =1$, \textit{i.\,e.} one He$^*$ is excited at an arbitrary position in the droplet. We assume that the He droplet has radius $R_D=63$~\AA, and the Li atom is at distance $R_\mathrm{Li}=R_D+d_\mathrm{Li}-d_d$ from the droplet center, where $d_\mathrm{Li}=5.9~$\AA~is the distance of the Li atom from the droplet surface, and $d_d=2.7~$\AA~is the depth of the dimple as given in Ref.~\cite{Hernando:2012}. The volume of the He droplet is $V_D=4\pi R_D^3/3$. For simplicity, we assume that the He density distribution has a sharp edge at distance $d_\mathrm{Li}$ from the Li atom. For this geometry, the fraction of the total rate of exchange ICD \textit{versus} the total rate of direct ICD (integral over $\Gamma_D^{^1\mathrm{S}}$ for $R>12$~\AA) is $5\times 10^{4}$. Thus, for all realistic configurations involving a Li atom attached to a He cluster or nanodroplet, ICD proceeds nearly exclusively by charge exchange despite the rather large interatomic distances. Most likely, this conclusion holds for all other alkali atoms attached to He droplets due to the similar structure of their valence state orbitals. Taking into account that in some cases ICD occurs at shorter He$^*$-Li distance than $d_\mathrm{Li}$ due to atomic motion preceding the ICD, the overall proportion of exchange ICD is even further enhanced. 

When neglecting atomic motion, the droplet-weighted average ICD rate results in characteristic times $t^{^1\mathrm{S}}_D=1/\Gamma_D^{^1\mathrm{S}}$ and  $t^{^3\mathrm{S}}_D=1/\Gamma_D^{^3\mathrm{S}}$ which fall in the $\mu s$ range, given the considerable contribution of large values of $R$ to the weighted average. Thus, in the real droplet system, only He atoms in the first layer of the dimple next to the Li atom at $R\sim 6$~\AA~effectively undergo ICD with decay times $t^{^1\mathrm{S}}=900$~ps and $t^{^3\mathrm{S}}=140$~ps. Already for those He atoms excited in the second layer of the dimple at $R\sim 9$~\AA, the ICD times given by the decay rate $\Gamma^{^1\mathrm{S},\,\,^3\mathrm{S}}$ are much longer, $t^{^1\mathrm{S},\,\,^3\mathrm{S}}>100$~ns, implying that massive changes of the local geometry take place prior to ICD, \textit{e.\,g.} due to the formation of a bubble around the He$^*$~\cite{MudrichRelaxation}, the migration of He$^*$ to the droplet surface, He$_2^*$ dimer formation, and other effects associated with atomic motion. 

\begin{figure}
	\center
	\includegraphics[width=0.6\columnwidth]{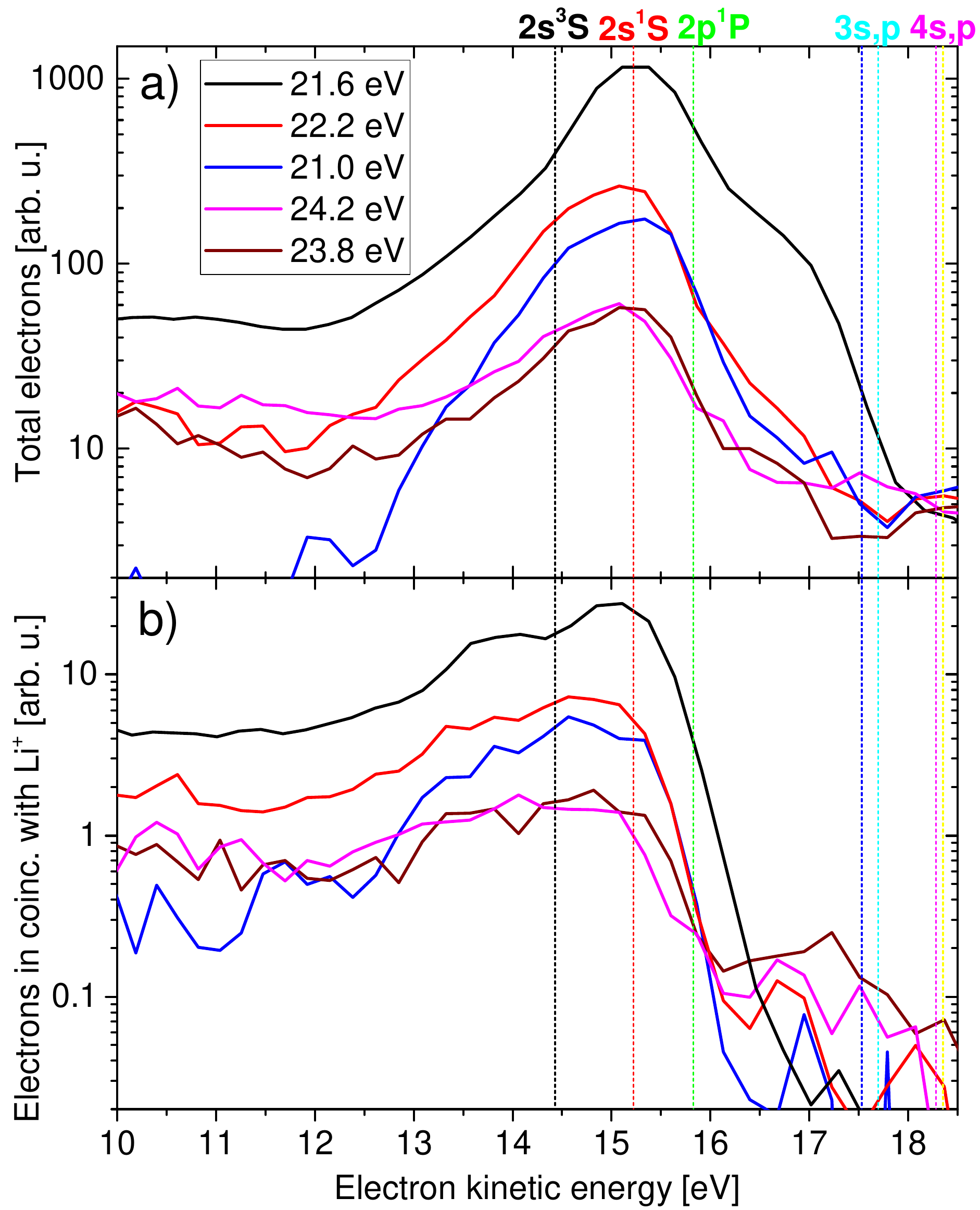}\caption{\label{fig4} Spectra in logarithmic scale of all electrons a) and of electrons recorded in coincidence with Li$^+$ ions b) for doped He nanodroplets resonantly excited at various photon energies.}
\end{figure}
Finally, we address the question how the ICD signals depend on the excited states of the He nanodroplet. Fig.~\ref{fig4} shows electron spectra recorded for Li-doped He nanodroplets at various photon energies $h\nu$ when all emitted electrons are measured (a), or only electron-Li$^+$ ion coincidences are detected (b). At $h\nu = 21.0$~eV (blue lines), the He droplets are excited directly to the 1s2s$^1$S droplet state, whereas at $h\nu = 21.6$ and $22.0$~eV (black and red lines), the 1s2p$^1$P droplet state is excited~\cite{Joppien:1993}. At $h\nu = 23.8$ and $24.2$~eV (brown and pink lines), higher states correlating to the 1s3p and 1s4p levels are reached. Most notably, at all values of $h\nu$, the electron spectra are dominated by ICD involving the 1s2s states. Only for $h\nu > 23$~eV, a contribution of the 1s3s,\,p and 1s4s,\,p states to the ICD signal of about $10$~\% of the corresponding signal from the 1s2s states is observed [Fig.~\ref{fig4} b)]. This indicates that internal droplet relaxation is much faster than ICD, \textit{i.~e.} the 1s2s state is populated in much less than $t^{^1\mathrm{S}}=900$~ps. This is in agreement with relaxation times $\lesssim 1$~ps measured directly using pure He nanodroplets~\cite{Ziemkiewicz:2015}. To directly map the dynamics of the various ICD channels, pump-probe experiments using tunable ultrashort EUV laser pulses would be highly desirable.

In summary, we have shown that for He nanodroplets doped with alkali metal atoms (Li and Rb), interatomic autoionization induced by resonant excitation of the He droplet is predominantly driven by charge exchange ICD (equivalently termed Penning ionization), even though it proceeds at large interatomic distances given by the initial configuration of the doped He nanodroplet. This is due to the diffuse structure of the electron orbitals of both the alkali metal atom and the excited He$^*$ atom. This case drastically differs from most systems where ICD was studied previously, \textit{e.~g.} by inner-shell excitation or ionization of rare gas dimers and clusters~\cite{Jahnke:2007,Jahnke:2015}. 
It is likely that charge exchange ICD is also the dominant autoionization mechanism in other systems involving valence-excited rare gas atoms, such as most Penning reactions~\cite{Siska:1993} as well as the recently studied autoionization of multiply excited rare gas clusters~\cite{Kuleff:2010,Ovcharenko:2014,Iablonskyi:2016,Serdobintsev:2018}. Furthermore, we find that nearly irrespective of the initial level of excitation of the He nanodroplet, autoionization occurs out of the 1s2s-correlated He levels due to ultrafast droplet relaxation. While direct ICD mostly involves the droplet perturbed 1s2s$^1$S state, charge exchange ICD occurs out of the 1s2s$^1$S and the 1s2s$^3$S states in close analogy to traditional Penning ionization of colliding atoms at thermal velocities. Besides giving insight into fundamental interatomic decay processes, the method of measuring coincidence electron and ion spectra for surface-bound atoms may be useful for probing the relaxation dynamics of other types of clusters and nanoparticles as well.

\section{Experimental methods}
The experiments are performed using a He nanodroplet apparatus combined with a velocity-map imaging photoelectron-photoion coincidence (VMI-PEPICO) detector at the GasPhase and CiPo beamlines of Elettra-Sincrotrone Trieste, Italy. The apparatus has been described in detail elsewhere~\cite{Buchta:2013,BuchtaJCP:2013}. Briefly, a beam of He nanodroplets is produced by continuously expanding pressurized He (50~bar) of high purity out of a cold nozzle (14~K) with a diameter of 5~$\mu$m into vacuum, resulting in a mean droplet size of $\bar{N}_\mathrm{He} = 2.3\times 10^4$ He atoms per droplet. Further downstream, the beam passes a mechanical beam chopper used for discriminating droplet-beam correlated signals from the background. The He droplets are doped with Li and Rb atoms by passing through two vapor cells containing elementary alkali metal heated to 400 $^\circ$C and 90 $^\circ$C, respectively.

In the detector chamber, the He droplet beam crosses the synchrotron beam in the center of the VMI-PEPICO detector at right angles. By detecting either electrons or ions with the VMI detector in coincidence with the corresponding particles of opposite charge on the TOF detector, we obtain either ion mass-correlated electron images or mass-selected ion images. Kinetic energy distributions of electrons and ions are obtained by Abel inversion of the images~\cite{Dick:2014}. The energy resolution of the electron spectra obtained in this way is $\Delta E/E=6$\%.


\begin{acknowledgement}
	M.M. and L.B.L. acknowledge financial support by Deutsche Forschungsgemeinschaft (DFG, German Research Foundation, projects MU 2347/10-1 and BE 6788/1-1:1) and by the Carlsberg Foundation. A.C.L gratefully acknowledges the support by Carl-Zeiss-Stiftung. S.R.K. thanks Max Planck Society and D.S.T., Govt. of India, for support. E.F. gratefully acknowledges support by the Villum Foundation. N.S. acknowledges financial support by Agence Nationale de la Recherche through the program ANR-16-CE29-0016-01. The research leading to this result has been supported by the project CALIPSOplus under Grant Agreement 730872 from the EU Framework Programme for Research and Innovation HORIZON 2020.	
\end{acknowledgement}


\begin{mcitethebibliography}{14}
\bibitem[Not()]{Note-1}
This is a note. The text will be moved the the references section. The title of
  the section will change to ``Notes and References''.\relax
\mciteBstWouldAddEndPunctfalse
\mciteSetBstMidEndSepPunct{\mcitedefaultmidpunct}
{}{\mcitedefaultseppunct}\relax
\EndOfBibitem
\end{mcitethebibliography}
\providecommand{\latin}[1]{#1}
\makeatletter
\providecommand{\doi}
  {\begingroup\let\do\@makeother\dospecials
  \catcode`\{=1 \catcode`\}=2 \doi@aux}
\providecommand{\doi@aux}[1]{\endgroup\texttt{#1}}
\makeatother
\providecommand*\mcitethebibliography{\thebibliography}
\csname @ifundefined\endcsname{endmcitethebibliography}
  {\let\endmcitethebibliography\endthebibliography}{}

\end{document}


\title{Dopant ion kinetic energy distribution}
	\date{}
	\maketitle
	\date{}

The difference between the total electron spectra and the electron spectra measured in coincidence with Li$^+$ and Rb$^+$ dopant ions indicates that  the ejection of the dopant ions from the He nanodroplets is facilitated by the atomic motion prior to ICD. This picture is supported by the distribution of dopant ion kinetic energy $K_i$ inferred from the ion velocity-map image recorded in electron-ion coincidence detection mode. The measured distribution (black line), shown in Fig.~S1, features two contributions, a sharp maximum centered around 0.1~eV, and a low-intensity tail reaching up to $K_i>1$~eV. The tail to higher energies is likely due to the dissociation of Li$_2$ dimers and other Li containing heterodimers forming on the He nanodroplets at low concentrations. The red line in Fig.~S1 shows a simulated ion spectrum inferred from the literature PIES~\cite{Ruf:1987} by linearly transforming its argument according to $K_i=\left(E_\mathrm{He}-E_i-K_e\right)m_\mathrm{He}/(m_\mathrm{He}+m_\mathrm{Li})$. Here, $E_\mathrm{He}$ is the atomic term values of the 1s2s$^1$S and $^3$S states, and $m_\mathrm{He}$ and $m_\mathrm{Li}$ are the atomic masses of He and Li. This corresponds to the assumption that any difference between the measured electron energy and the values for the atomic levels is converted into kinetic energy of the moving He and Li atoms. The large peak at 0.1~eV is the contribution of the 1s2s$^1$S state undergoing ICD. It reproduces the narrow peak in the experimental distribution rather well. The second peak at 0.27~eV is due to the 1s2s$^3$S state whose contribution is 0.65 times that of the 1s2s$^1$S state, as inferred from the fitted simulated electron spectrum shown by the light blue line in Fig.~S1 b). This peak is not directly seen in the experimental spectrum. We assume that it is spread out towards large energies and adds to the extended tail in the range $K_i>0.2$~eV. Presumably the kinematics of the He-Li$^+$ dissociation deviates from the assumed diatomic system due to the presence of more He atoms nearby, as previously observed for laser-excited alkali metal atoms detaching from He nanodroplets~\cite{Hernando:2012,Vangerow:2014}. An accurate modeling of the dynamics ensuing dopant ICD would require a many-body simulation of the full process including the correlated electron dynamics and the nuclear motion.

\begin{figure}[H]
	\center
	\includegraphics[width=0.6\columnwidth]{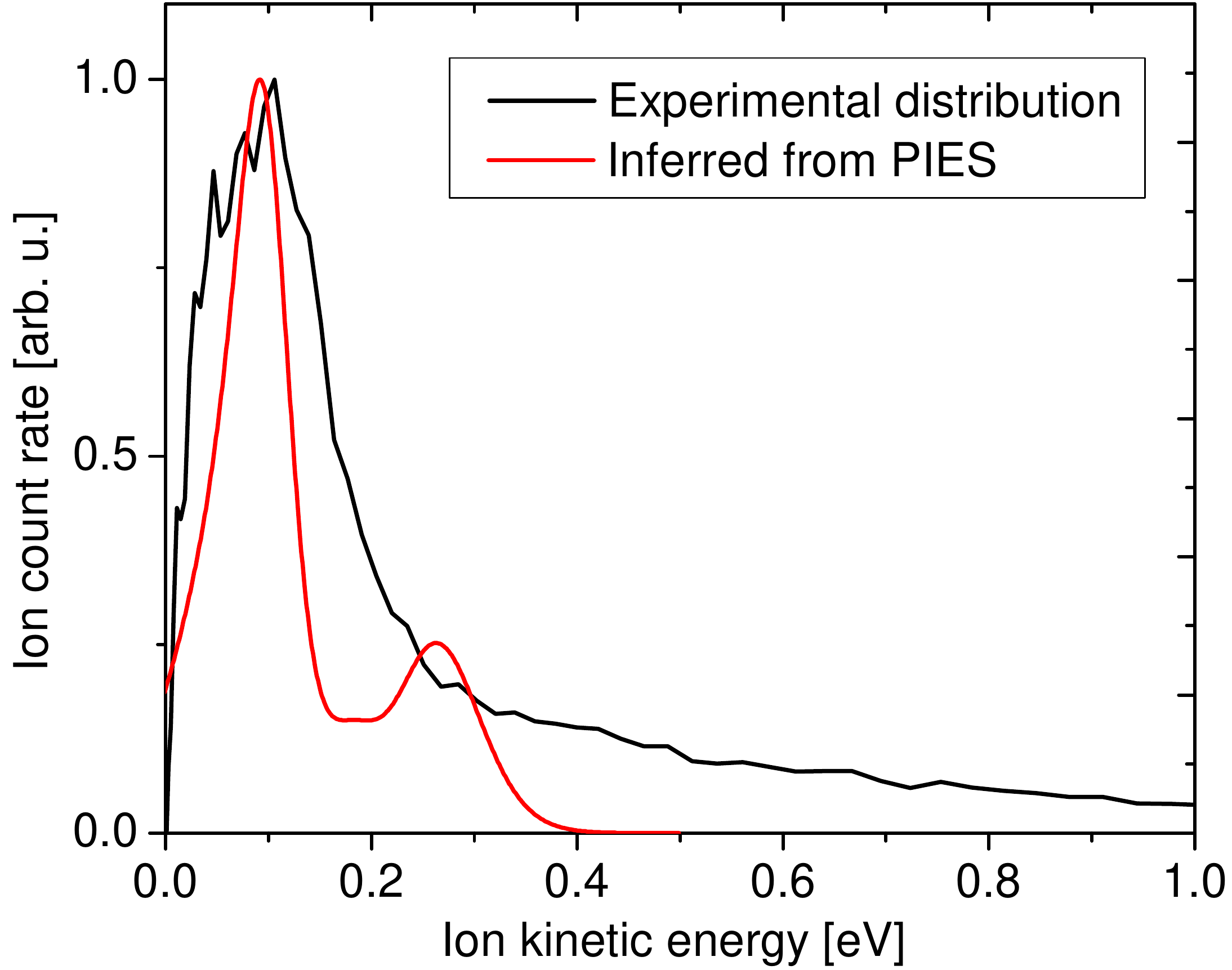}\caption{\label{fig3} Kinetic energy distribution of Li$^+$ ions created by ICD on He droplets. The red line depicts the Li$^+$ kinetic energy distribution calculated from the collisional Penning ionization electron spectrum~\cite{Ruf:1987}. }
\end{figure}